# Atomic and electronic structure of nitrogen- and boron-doped phosphorene

Danil W. Boukhvalov[1,2]

[1]*Department of Chemistry, Hanyang University, 17 Haengdang-dong, Seongdong-gu, Seoul 133-791, Korea*

[2]*Theoretical Physics and Applied Mathematics Department, Ural Federal University, Mira Street 19, 620002 Ekaterinburg, Russia*

*First principle modeling of nitrogen- and boron-doped phosphorene demonstrates the tendency toward formation of highly ordered structures. Nitrogen doping leads to the formation of –N-P-P-P-N- lines. Further transformation to -P–N-P-N- lines across the chains of phosphorene occurs with increasing band gap and increasing nitrogen concentration, which coincides with the decreasing chemical activity of N-doped phosphorene. In contrast to the case of nitrogen, boron atoms prefer to form -B-B- pairs with the further formation of -P-P-B-B-P-P- patterns along the phosphorene chains. The low concentration of boron dopants converts the phosphorene from a semiconductor into a semimetal with the simultaneous enhancement of its chemical activity. Co-doping of phosphorene by both boron and nitrogen starts from the formation of -B–N- pairs, which provide flat bands and the further transformation of these pairs to hexagonal BN lines and ribbons across the phosphorene chains.*

E-mail: danil@hanyang.ac.kr

## 1. Introduction

Two-dimensional systems beyond graphene have attracted experimental and theoretical attention in recent years. [1-3] These materials are discussed as a possible material for photonics, electronics, catalysis and the building blocks of "sandwich" structures. [4] Phosphorene is a 2D material that also has great potential. [5] High carrier mobility and other attractive electronic properties make it a possible candidate for employment in 2D-based devices. [6] One issue in the further application of phosphorene is its chemical stability under ambient conditions, which has been recently discussed both experimentally [7-9] and theoretically. [10,11]

Recent experimental [12] and theoretical [13,14] works discuss valuable changes in the electronic structure and chemical properties of boron- and nitrogen-doped graphene. Nitrogen-doped graphene has been discussed as a possible catalyst for the oxygen reduction reaction. [15-17] Because there is large number of possible combinations of N–C bonds caused by the mismatch of nitrogen and carbon valences and comparable C–N and C-C bond lengths, it is very difficult to obtain the atomic structure of nitrogen-doped graphene from first principle calculations. [13,14] Experiments have also demonstrated significant diversity in the atomic structure of nitrogen-doped graphene, which depend on the combination of several experimental conditions during the fabrication of samples. [18,19] In contrast to graphene, the atomic structure of phosphorene is anisotropic and can be discussed as the summation of one-dimensional chemical bonded chains (see Fig. 1) which will discussed further in the text as the lines of phosphorene. The valences of P, N and B coincide, and typical P-N and P-B distances (1.6 and 1.8 Å, respectively) [20,21] are rather different from the P–P distance in phosphorene (approximately 2.2 Å, see Table I). The presented causes strictly limit the number of possible configurations of the impurities and make possible only the substitution of phosphorous atoms in contrast to formation of local $C_xN_y$ structures in N-doped graphene and make it possible to discuss the atomic structure of N- and B-doped phosphorene obtained from first principles.

In this work, we explore the atomic and electronic structure of 1.4-37.5% nitrogen and boron dopants in phosphorene by the step-by-step increase of impurity atoms in a supercell. For single impurities and pairs, we examine the energetics of hydrogenation by molecular hydrogen considering all possible configurations of adatoms on the phosphorene surface.

## 2. Computational model and method

For the modeling, a rectangular supercell of phosphorene with 72 phosphorous atoms was used and performed step-by-step increasing of substitutional defects (see Fig. 1). For all steps, all possible positions of the added impurity were checked and the configuration with the lowest total energy of the system was used. Because the calculation of the formation energies requires knowledge of the initial compounds, which was used as a source of the dopant in this case, it cannot be calculated. To study the energetics of the doping process, the difference between the total energy of the impurity atom at the current and initial step of doping was calculated using the formula

$\Delta E = (E_{final} - (E_{init} + nE_i))/n$,

where $E_{final}$ is the total energy of the doped phosphorene with N+n impurity atoms, $E_{init}$ with N impurity atoms, and

$E_i = E_{single} - E_{pure}$,

where $E_{single}$ is the energy of the supercell with the first impurity, and $E_{pure}$ – energy of the same supercell of pristine phosphorene. The value of $\Delta E$ demonstrates how energetically favorable the N+n$^{th}$ impurity is in the current position compared to another single remote impurity. The calculation of the hydrogen chemisorption energy was performed by

$E_{chem} = E_{ph2H} - E_{ph+H2}$,

where $E_{ph+H2}$ is the energy of phosphorene with the $H_2$ molecule physisorbed on the reaction site and $E_{ph2H}$ is the energy of phosphorene with two chemisorbed hydrogen atoms.

Pseudo-potential code SIESTA [22] was used to perform energy calculations of the various atomic structures of functionalized phosphorene with the density functional theory (DFT). All calculations were performed by the local density approximation (LDA) [23] with spin polarization. To model the phosphorene monolayer, a rectangular supercell with 72 phosphorus atoms was used (see Fig. 1). The atomic positions were fully optimized. During optimization, the ion cores were described by norm-conserving, non-relativistic pseudo-potentials [24] with cut-off radii of 1.85, 1.20, 1.15 and 1.25 a.u. for P, B, N and H, respectively. The wave functions were expanded with a double-$\zeta$ plus polarization basis of the localized orbitals for all atoms except hydrogen. Double-$\zeta$ was used for hydrogen. The force and total energy was optimized with an accuracy of 0.04 eV/Å and 1 meV, respectively. All calculations were performed with an energy mesh cut-off of 360 Ry and a **k**-point mesh of 12×10×1 in the Monkhorst-Pack scheme. [25]

## 3. Nitrogen-doped phosphorene

A nitrogen atom has an orbital occupancy similar to a phosphorous atom ($2s^2 2s^3$ and $3s^2 3s^3$ respectively). This similarity leads to nonsignificant changes in the electronic structure of the phosphorene with a single nitrogen impurity (Fig. 2a). The nearly three-times-smaller ionic radius of nitrogen (0.16 Å vs 0.44 Å for phosphorous) provides changes bonds with the nearest phosphorous atoms on the order of 0.5 Å (See Table I) and a valuable deviation of nitrogen impurity from its stoichiometric position (Fig. 1a). These changes in the atomic structure cause changes in the overlaps between phosphorous and nitrogen orbitals and especially between sigma orbitals with lone pairs of electrons (see discussion in Ref. [11]), which leads to light doping of the phosphorene matrix from nitrogen (Table I).

For evaluation of the effect of nitrogen doping on chemical activity, the decomposition of molecular hydrogen over three possible sites on the phosphorene surface was calculated (Fig. 3). The calculations (Table II) demonstrate that despite light doping of phosphorene from nitrogen, chemisorption of the pair of hydrogen atoms is less favorable than pristine phosphorene. The cause of this decay in chemical activity is the lattice distortions caused by nitrogen doping. Chemisorption of adatoms provides a light shift down from the plane of phosphorous atoms that should provide increasing P-N bonds, but in nitrogen-doped phosphorene these distances are already larger than the typical values of these bonds (1.8 Å vs 1.6 Å) [20] and its further magnification is very energetically favorable. Therefore, a small amount of nitrogen impurities can significantly decrease the chemical activity of phosphorene without significant changes to the electronic structure, and nitrogen-doped phosphorene should be safer under ambient conditions than pure phosphorene. [7, 11]

The next step of our survey was to check the case of increasing nitrogen concentration. To search the place next to the impurity atom, calculations were performed for various possible positions and it was determined that the lowest energy corresponds with the placement of the second nitrogen atom in the next line of phosphorene exactly opposite the first impurity (Fig. 1b). To understand the cause of the disposition of the second impurity and the energetic favorability (Fig. 4a) of the formation of the –N-P-P-P-N- chain, the effect of the nitrogen impurities on the lattice parameters of the supercell needs to be confirmed. The first substitution impurity provides compression of the phosphorene lattice along the chains and small expanding of lattice across the chains (Fig. 4b,c) and the presence of the second impurity obeys this pattern. The described change of the geometry of the whole supercell coincides with the local atomic structure of the nitrogen pair (Table I) when the N-P distances and changes of the Mulliken populations are nearly the same as in the case of the single impurity. The similarity of the case of the nitrogen pair with a single nitrogen impurity also leads to similarity in the chemical activity (Table II). The presence of the second impurity slightly decreases the chemisorption energies but remains greater than in the case of pristine phosphorene.

The last step of the survey of nitrogen doping was step-by-step increase of the concentration up to 30%. The results of the calculations demonstrate that there will be the formation of the first –N-P-P-P-N- chains (Fig. 1c) and then further transformation to wider –N-P-N-P- chains (Fig. 1d) with decreasing lattice parameters. A further increase of the nitrogen concentration up to nearly 40% turns the phosphorene to a phosphorene-like NP compound. If the presence of a small amount of nitrogen does not significantly affect the electronic structure of phosphorene (Fig. 2a,b) then further increase of nitrogen content provides visible increasing of the band gap (Fig. 2c,d). Taking into account underestimating the band gap value within the standard DFT, the increasing difference between calculated within DFT values, and more exact calculations with the employment GW method, [26] the bandgap can be estimated for a single nitrogen impurity (Fig. 1a) as approximately 1.5 eV and for a single –N-P-N-P- chain (Fig. 1d) as 3 eV.

**4. Boron-doped phosphorene**

A single boron impurity deviates from the stoichiometric position in phosphorene similar to the case of the single nitrogen impurity (Fig. 5a). However, the difference in the occupancies of the orbitals provides nearly an opposite effect to the electronic structure and chemical properties. In contrast to nitrogen with five elections in the second shell, boron has only three electrons. Three electrons participate in the formation of σ-bonds, and one σ orbital, which contains a lone pair of electrons in the case of phosphorous and nitrogen, is now empty and attract electrons from the orbitals of the nearest phosphorous atoms (see Table I). This visible redistribution of the charge between the matrix and dopant provides changes in the electronic structure: phosphorene turns from a semiconductor to semimetal (Fig. 6a). This effect is similar to recently discussed semiconductor-metal transition in metal-doped phosphorene. [27] The lengths of the P-B bonds in boron-doped phosphorene (Table I) are close to the values measured for various compounds (1.84-1.96 Å). [21] The combination of the significant charge redistributions and P-B distances comparable to typical values makes decomposition of the molecular hydrogen over boron centers energetically favorable.

The next step is a survey to determine the position of the second boron impurity in the supercell. Different positions of the second impurity were examined and the most energetically favorable position was nearest the first impurity but in another plane of phosphorene as shown in Fig. 4b. The difference with nitrogen doping has two causes: (i) minimization of the in-plane distortions (see Fig. 4b,c) and (ii) possibility of the B-B bonds in the solid phase. In contrast with nitrogen, boron at ambient conditions forms a large number of various stable solid allotropes, which makes it possible to minimize the local distortion of the atomic lattice via the formation of a boron pair in an area already distorted by the first impurity. The formation of strong B-B bonds enhances the metallization of doped phosphorene (Fig. 6b) and decays the charge transfer from the phosphorene matrix to boron impurities (Table II), which, with local distortions, provides decreasing chemical activity of B-doped phosphorene (Table II).

Similar to the case of the second nitrogen impurity, the formation of the pair of boron impurities is slightly more energetically favorable than single impurities. In contrast to the N-doped phosphorene, the next steps of increasing the concentration of impurities do not provide a significant energy gain. The cause of this is the greater changes to the lattice parameters of the doped phosphorene (Fig. 4b,c) that correspond to distortions of the whole supercell by increasing distortion propagation. A more energetically favorable scenario for the case of increasing boron concentration is the formation of –P-P-B-B-P-P- patterns (Fig. 5c). The formation provides segregation of the boron-doped area from the undoped area of the phosphorene matrix and as result, B-doped phosphorene turns from a semimetal to the semiconductor (Fig. 6c). A further increase of the boron concentration provides a significant (Fig. 4b,c) and very energetically favorable (Fig. 4a) distortion of the phosphorene matrix with deviation from the stoichiometric positions. The cause of this effect is the combination of the phosphorene matrix distortions caused by boron doping with the tendency of the rotation of B-B pairs (see side views on Fig. 5). At low concentrations, boron pairs cannot realize this tendency completely, but when the next phosphorene line appears, another B-B pair lattice distortion occurs that provides additional distortions and further turns the doped phosphorene to an amorphous-like form. Thus, the maximum amount of boron doping to phosphorene is 11%, which corresponds with the alteration of –B-B-P-P- and pure phosphorene lines. The energetics of the increasing boron amount (Fig. 4a) demonstrate that the first step after finishing the formation of uniform –B-B-P-P- lines (11.1%) is rather energetically unfavorable, and creation of reported boron-rich chains can be described as self-limited.

## 5. Boron and nitrogen co-doping

The last step of the survey is the simultaneous modeling of doping of phosphorene by nitrogen and boron. First, the position of a single boron impurity is checked in phosphorene with a single nitrogen impurity. All possible positions of the boron impurity were checked and it was determined that the most energetically favorable location corresponds to the substitution nearest to the nitrogen impurity phosphorous atom in the other plane (Fig. 7a). The nature of this B-N pairing can be explained by redistribution of the charges (Table I). There was significant charge transfer from the nitrogen to boron, suggesting that the formation of a covalent-polar chemical bond between these impurities was similar to the case of pure boron nitride. The formation of this bond leads to the decreasing chemical activity of the boron impurity (Table II) because empty orbitals became partially saturated. Similarity with boron nitride can also be observed in the electronic structure, where the band gap of the pair is wider than in the phosphorene matrix (Fig. 8). Another feature of the electronic structure of the B-N pairs of phosphorene at low concentrations is the appearance of flat bands (see insets on Fig. 8a,b) that encourage the synthesis of boron and nitrogen co-doped phosphorene for superconductivity, which in 2D systems is caused by flat bands. [28] One more difference between the B-N pair and B-B is the more valuable deviation of impurity both (B and N) impurity atoms from the stoichiometric position, which can be explained in terms of the formation of the pair of atoms in sp2 hybridization that is corresponding with planar configurations inside the non-planar matrix of sp3 hybridized neighbors.

The position of the second and next B-N pairs were examined, and it was determined that they obey the same tendencies that were obtained for the nitrogen-doped phosphorene. First, this pair forms -B-N-P-P- patterns across the lines of phosphorene with further substitutions of -P-P- pairs next to -B-N- pairs and the formation of BN-lines (Fig. 7b). Increasing B–N content provides increases of the band gap of whole boron and nitrogen-doped phosphorene (Fig. 8c). In contrast to the formation of –B-B– chains (Fig. 5c), the formation of boron nitride chains is very energetically favorable (Fig. 4a) for two reasons: the energy gain from the formation of covalent-polar B-N bonds and the minimization of local geometry distortions by the formation of nearly planar B-N chains. This chain formation also provides smaller distortions of the whole phosphorene matrix due to separation of impurity patterns that makes the formation of these BN chains rather attractive. The next B–N pairs prefer to form the next line with further formation of planar BN nanoribbons within the phosphorene matrix (Fig. 7c).

## 6. Conclusions

Modeling demonstrates that despite the same valence behavior of boron and nitrogen in the phosphorene matrix and its influence on electronic and chemical properties, the compound formed varies considerably. Nitrogen prefers to form first -N-P-P-P-N- lines across lines of phosphorene. A further increase in nitrogen content provides the formation of large NP ribbons. Nitrogen doping also provides an increasing band gap and energy cost of the molecular hydrogen decomposition. Boron doping provides the lowest concentration of impurities to turn phosphorene from a semiconductor to semimetal and significantly increases the chemical activity of phosphorene. This contrasts with the case of nitrogen doping caused by the presence of empty orbitals on the boron 2p shell, which attracts some electron density from the 3p orbitals of phosphorous, and the values of the P-B distances in phosphorene are comparable to others, in contrast to the P-N bonds in phosphorene that are shorter. Further increases in the boron content leads to the formation of -B-B-N-N- lines along the phosphorene chains, which become amorphous with increasing boron content. The combination of boron and nitrogen dopants leads to the first formation of -B-N- pairs with further transformation to boron nitride nanoribbons across the phosphorene chains. The coincidence of the valence of the host and guest atoms and anisotropy of the crystal structure of phosphorene enables formation of the ordered structure of nitrogen and boron impurities, in contrast to the case of the same dopants in graphene, and facilitates manipulation of the electronic and chemical properties of phosphorene by varying the type and concentration of the impurities.

**Acknowledgements** The work is supported by the Ministry of Education and Science of the Russian Federation, Project N 16.1751.2014/K

**Table I.** Interatomic distances (in Å) and changes of Mulliken populations (in electrons) for various configurations of dopants atoms in phosphorene

| Defect | In-plane | Inter-plane | Δe⁻ |
|---|---|---|---|
| pristine | 2.25 | 2.32 | - |
| N | 1.80 | 1.82 | -0.07 |
| 2N | 1.79 | 1.83 | -0.07 |
| B | 1.96 | 1.92 | 0.35 |
| 2B | 1.97 | 1.66 | 0.24 |
| B+N |  | 1.44 |  |
| N-site | 1.81 |  | -0.323 (-0.69) |
| B-site | 1.96 |  | 0.47 (0.55) |

**Table II.** Chemisorption energies (eV) of molecular hydrogen decomposition over various possible sites (see Fig. 3) of pure and doped phosphorene

| Defect | Along | Across | Bridge |
|---|---|---|---|
| pristine | -1.660 | +0.476 | +0.066 |
| N | -0.072 | -0.062 | +0.399 |
| 2N | -0.150 | -0.211 | +0.097 |
| B | -1.153 | -1.468 | -1.143 |
| 2B | -0.577 | -0.969 | -0.981 |
| B+N |  |  |  |
| N-site | +1.148 | +0.085 | -0.072 |
| B-site | -0.061 | -0.959 | -0.634 |

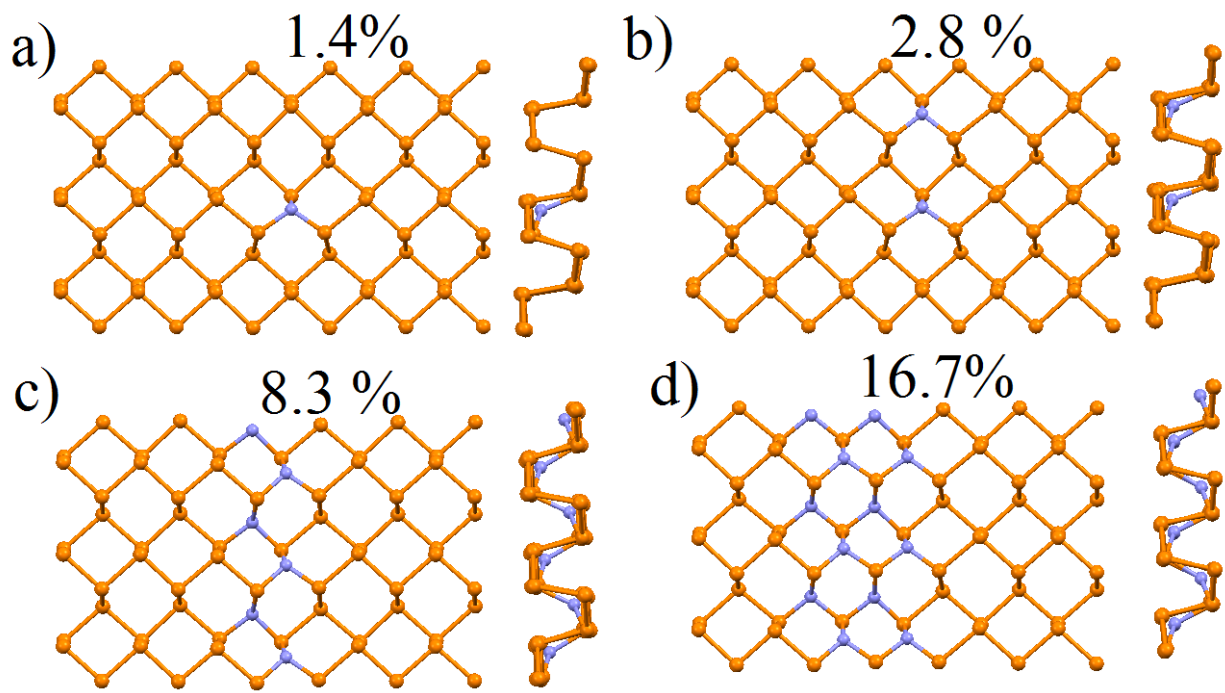

**Figure 1.** Top and side views of the optimized atomic structure of a phosphorene supercell with varying amounts of nitrogen impurities.

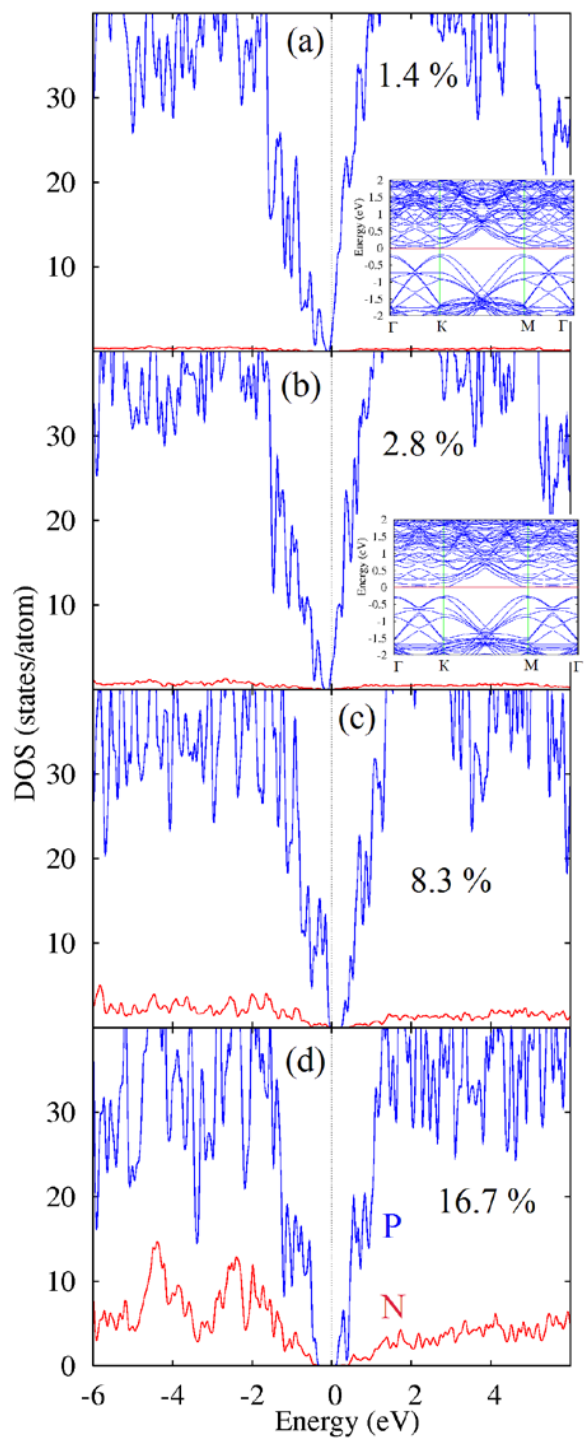

**Figure 2.** Density of states of phosphorous and nitrogen for various amount of impurities. Insets: band structure for selected configurations.

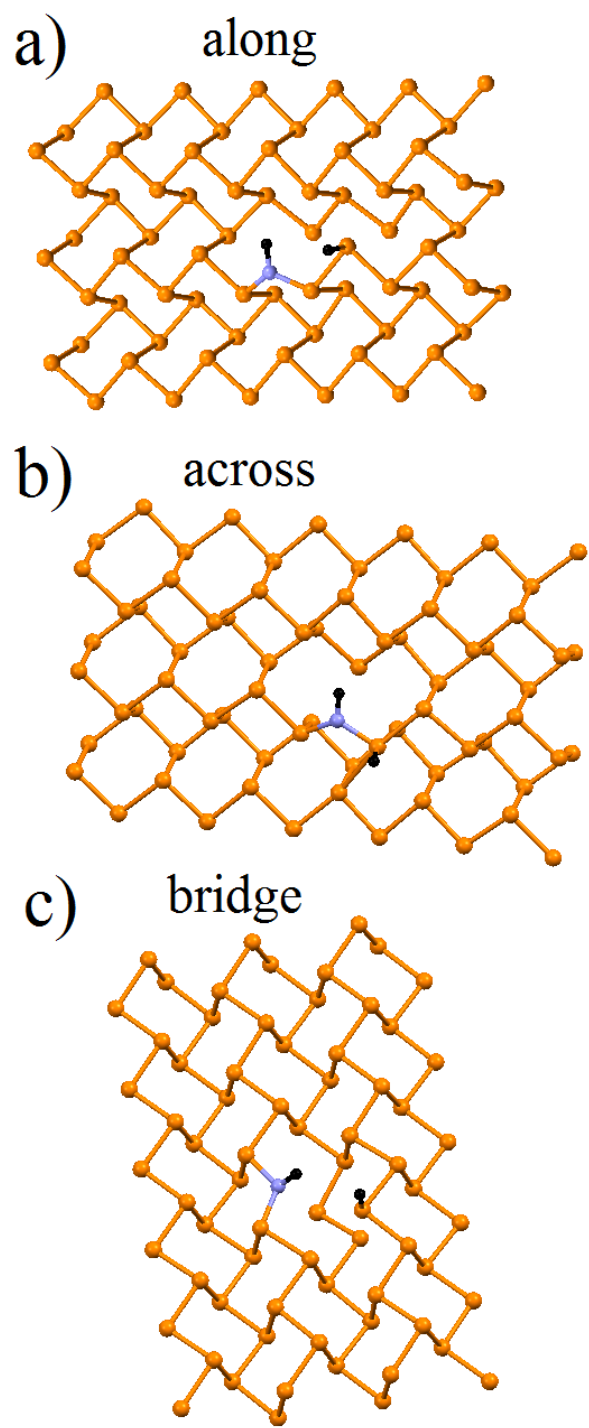

**Figure 3.** Optimized atomic structure of the results of molecular hydrogen decomposition over three possible sites of nitrogen-doped phosphorene.

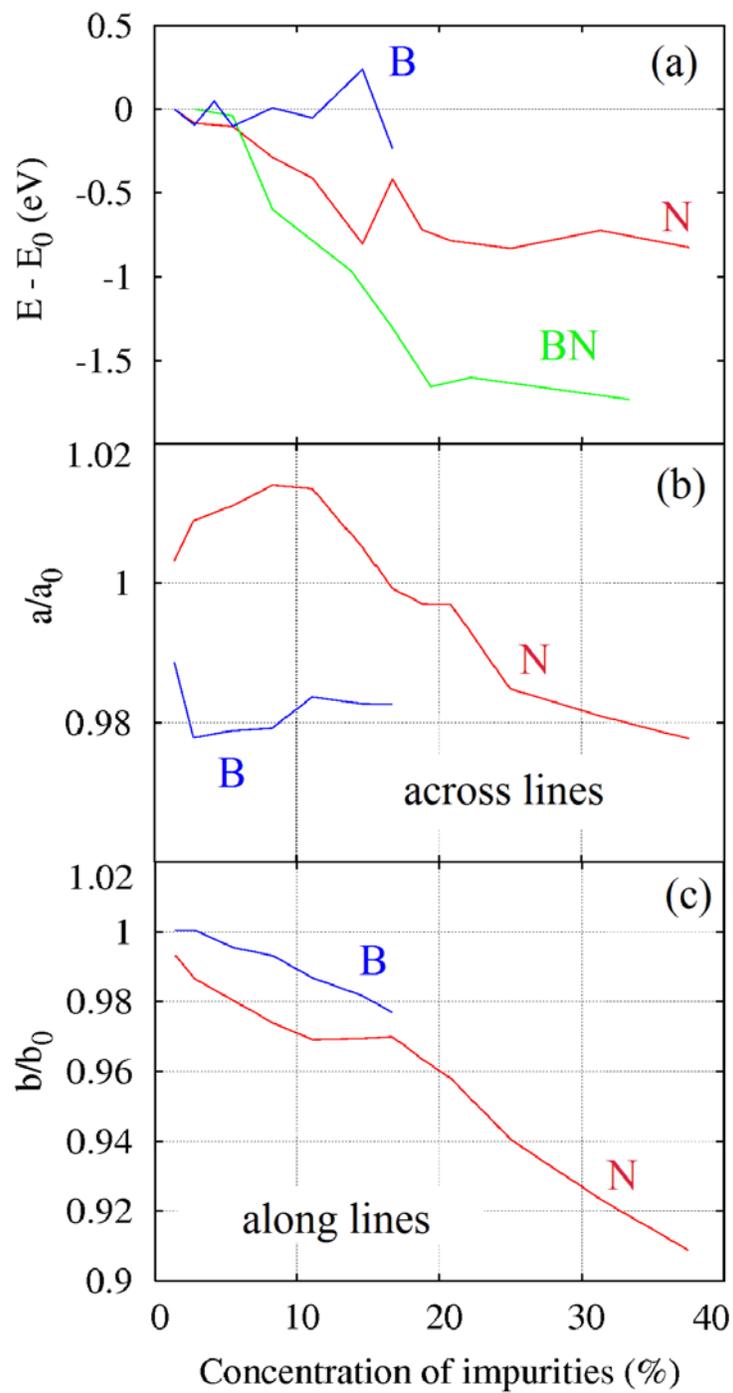

**Figure 4.** Energy differences (a) between the total energy of the impurity atom at current and lowest concentrations (for more details see Chapter 2) and distortion of the phosphorene supercell across (b) and along (c) lines as a function of the concentration of impurities. Note the boron-doping data is provided only below the change in doped phosphorene to an amorphous-like structure at 16.7% boron (d).

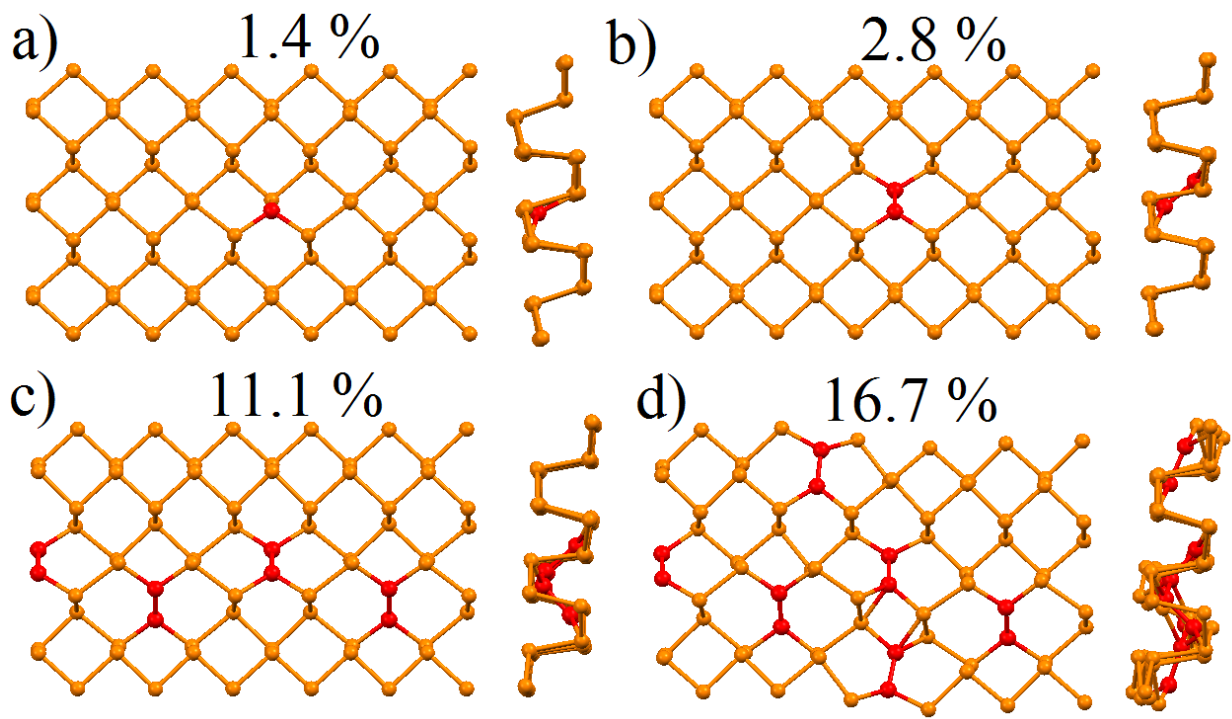

**Figure 5.** Top and side views of the optimized atomic structure of the phosphorene supercell with varying amounts of boron impurities.

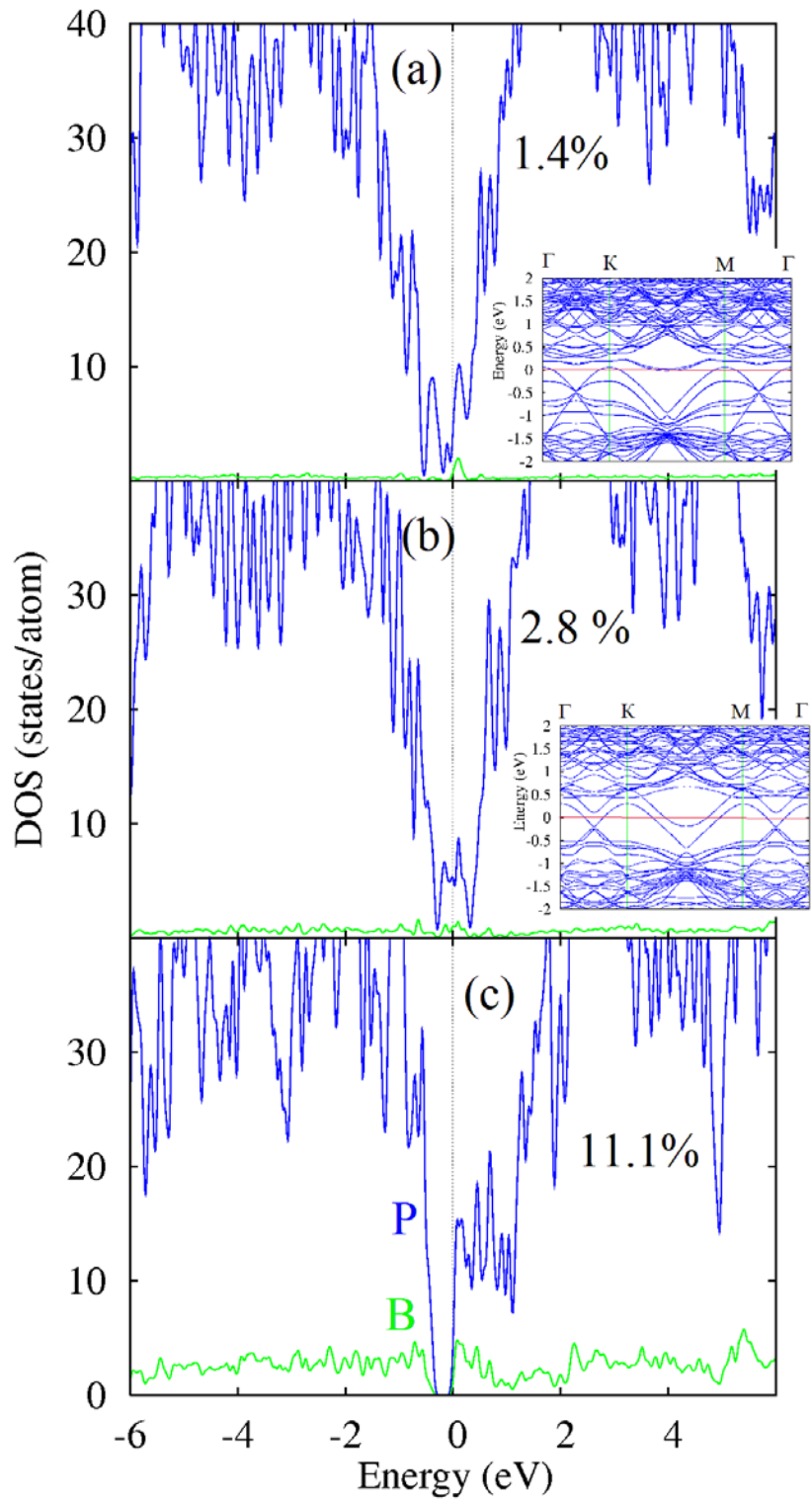

**Figure 6.** Density of the states of phosphorous and boron for varying amounts of impurities. Insets: band structure for selected configurations.

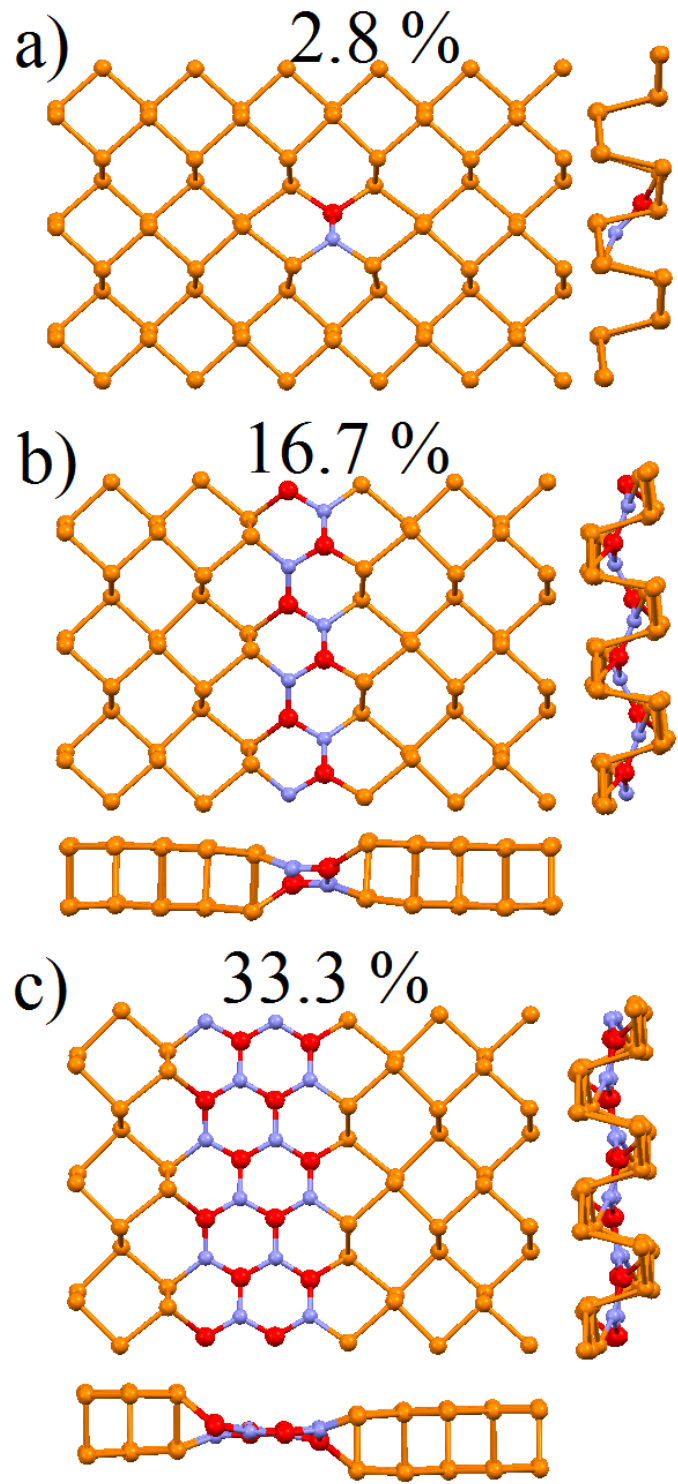

**Figure 7.** Top and side views of the optimized atomic structure of the phosphorene supercell with varying amounts of boron and nitrogen impurities.

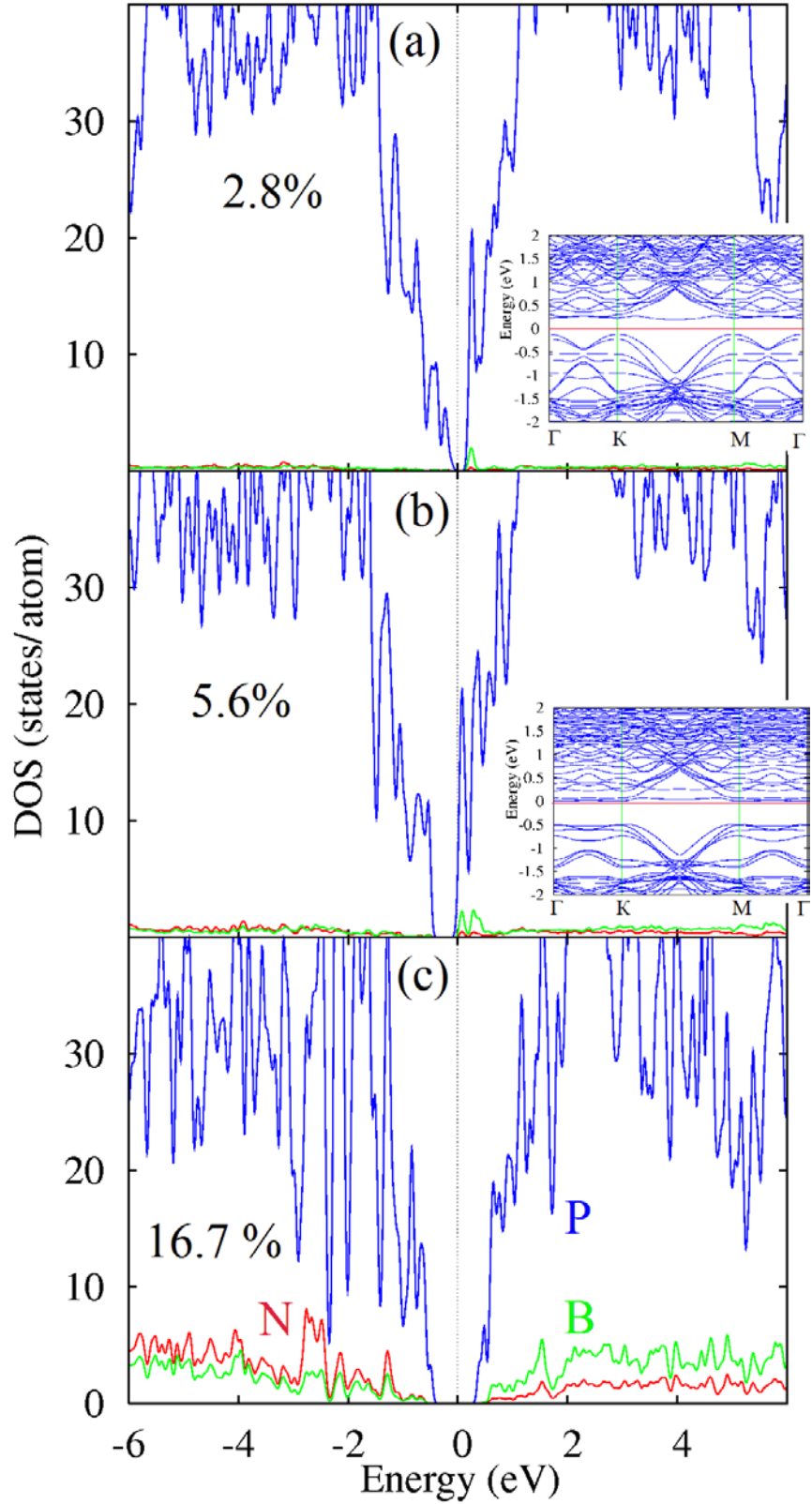

**Figure 8.** Density of states of phosphorous, boron and nitrogen for varying amounts of impurities. Insets: band structure for selected configurations.